\documentclass[reprint,showpacs,twocolumn,preprintnumbers,amsmath,amssymb,aip,graphicx,apl]{revtex4-1}

\usepackage{graphicx}
\usepackage{dcolumn}
\usepackage{bm}
\usepackage[mathlines]{lineno}
\usepackage{ulem}
\usepackage{color}

\begin{document}
\preprint{} 

\title[Short Title]{Coherent phonon study of (GeTe)$_{l}$(Sb$_{2}$Te$_{3}$)$_{m}$ interfacial phase change memory materials}

\author{Kotaro Makino}
\email{k-makino@aist.go.jp}
\author{Yuta Saito}
\author{Paul Fons}
\author{Alexander V. Kolobov}
\author{Takashi Nakano}
\author{Junji Tominaga}
\affiliation{Nanoelectronics Research Institute, National Institute of Advanced Industrial Science and Technology (AIST), Tsukuba Central 4, 1-1-1 Higashi, Tsukuba 305-8562, Japan}
\author{Muneaki Hase}
\affiliation{Institute of Applied Physics, University of Tsukuba, 1-1-1 Tennodai, Tsukuba 305-8573, Japan}

\date{\today}

\begin{abstract}
The time-resolved reflectivity measurements were carried out on the interfacial phase change memory (iPCM) materials ([(GeTe)$_{2}$(Sb$_{2}$Te$_{3}$)$_{4}$]$_{8}$ and [(GeTe)$_{2}$(Sb$_{2}$Te$_{3}$)$_{1}$]$_{20}$) as well as conventional Ge$_{2}$Sb$_{2}$Te$_{5}$ alloy at room temperature and above the RESET-SET phase transition temperature.
In the high-temperature phase, coherent phonons were clearly observed in the iPCM samples while drastic attenuation of coherent phonons was induced in the alloy.
This difference strongly suggests the atomic rearrangement during the phase transition in iPCMs is much smaller than that in the alloy.
These results are consistent with the unique phase transition model in which a quasi-one-dimensional displacement of Ge atoms occurs for iPCMs and a conventional amorphous-crystalline phase transition takes place for the alloy.

\end{abstract}

\maketitle
Phase change memory (PCM) technology is now growing in importance and is being utilized for applications in nonvolatile electrical memories as well as in optical disks.\cite{Wuttig07}
Pseudobinary chalcogenide alloys along the GeTe-Sb$_{2}$Te$_{3}$ tie-line such as Ge$_{2}$Sb$_{2}$Te$_{5}$ (GST) are commonly used because they exhibit large changes in optical and electrical properties between amorphous and crystalline phases.
Superlattice-like structured GST, sometimes referred to as interfacial phase change memory (iPCM), consisting of GeTe and Sb$_{2}$Te$_{3}$ sub-layers has received considerable attention since the SET-RESET phase switching energy in iPCM was demonstrated to be far smaller than that in conventional GST alloys.\cite{Simpson11}
Additionally, huge magnetoresistance \cite{Tominaga11} and anomalous magneto-optical Kerr-rotation \cite{Bang14} stemming from the topological nature of the iPCM structure \cite{Tominaga13} has been reported.
The combination of electrical memory switching and the unusual topological nature of iPCM is expected to pave the way to a new class of memory technology.

iPCM films have the structure of [(GeTe)$_{l}$(Sb$_{2}$Te$_{3}$)$_{m}$]$_{n}$, where $l$ and $m$ indicate the number of monolayers of each respective material, and $n$ is a integer that represents the number of combined sub-layers.
The phase transition is induced by a nanosecond electrical pulse.
Figure 1 (a) shows the primitive cell models for the high-resistance RESET and low-resistance SET states for the iPCM structure.
In iPCM, the RESET-SET operation is thought to be governed by a quasi one-dimensional transition of Ge atoms at the interfaces between tetrahedrally-coordinated sites and octahedrally-coordinated sites without melting and it is thereby considered that this mechanism should be different from that in conventional GST alloys where an amorphous-crystalline phase transition plays a dominant role. \cite{Wuttig07}
Spectroscopic measurements of lattice vibration therefore can serve as a powerful tool to elucidate the difference in phase transition mechanisms between iPCM and conventional GST alloys.

Herein, we report on optical pump-probe coherent phonon spectroscopy performed on iPCMs with a femtosecond laser to obtain a new perspective on the thermally-induced phase transition mechanism as indicated by the observed differences in the coherent phonons associated with the phase transition.
Ultrafast laser pulses absorbed in an opaque material excite Raman active coherent phonons via impulsive stimulated Raman scattering processes \cite{Merlin97} or abrupt changes in lattice potentials.\cite{Zeiger92}
Coherent phonons can be detected by time-resolved optical reflectivity measurement as change in reflectivity with the time periods of coherent phonons.
Coherent phonon spectroscopy provides good sensitivity for low frequency optical phonon modes ($<$ 200 cm$^{-1} $) which are dominant in GST materials \cite{Sosso09} and has been applied so far to GST materials to unveil their lattice dynamics, which is important for the understanding of the phase transition process.\cite{Forst00, Hase09, Makino11, Hernandez-Rueda11, Makino12, Shalini13}
For example, the amorphous GST alloy transforms into the fcc crystalline state or the hcp crystalline state depending on the annealing temperature \cite{Siegrist11}. It was found that a drastic change in coherent phonon spectra is induced by a thermally-induced phase transition.\cite{Forst00}

We investigated two different iPCM structures.
The first was [(GeTe)$_{2}$(Sb$_{2}$Te$_{3}$)$_{4}$]$_{8}$ (hereafter referred to as \{2,4\}-iPCM) and the second was [(GeTe)$_{2}$(Sb$_{2}$Te$_{3}$)$_{1}$]$_{20}$ (\{2,1\}-iPCM).
Both films were deposited on Si (100) substrates using a helicon-wave sputtering system.
Note that the layer thickness and, more importantly, the structures of the iPCM samples currently investigated are different from those of previously investigated samples by coherent phonon spectroscopy.\cite{Hase09, Makino11, Makino12}
The \{2,4\}-PCM sample is composed of 1.0 nm GeTe sub-layers and 4.0 nm Sb$_{2}$Te$_{3}$ sub-layers and the \{2,1\}-iPCM sample is consists of 1.0 nm GeTe sub-layers and 1.0 nm Sb$_{2}$Te$_{3}$ sub-layers on a 5 nm-thick Sb$_{2}$Te$_{3}$ growth controlling layer.\cite{Simpson12}
The total film thicknesses of two iPCMs structure were nearly identical.
The as-deposited amorphous GST alloy film that was used for comparison was fabricated by co-sputtering of GeTe and Sb$_{2}$Te$_{3}$ on a Si substrate using RF magnetron sputtering.
A 20-nm thick optically transparent ZnS-SiO$_{2}$ layer was deposited on top of all samples to prevent oxidation.
In the iPCM samples , the phase transition temperature from the RESET to SET state is $ \simeq $ 170 $^\circ $C and the electrical resistance reversibly changes with temperature. \cite{Bang14}
The alloy shows irreversible phase change, namely, the amorphous GST alloy transforms into a fcc phase at $ \simeq $ 150 $^\circ $C which further transforms into a hcp phase at $ \simeq $ 210 $^\circ $C.\cite{Forst00}
After crystallization, subsequent cooling of the alloy does not affect its phase as evidenced by the results of a temperature-dependent resistance measurement.\cite{Siegrist11} 
Thus the crystalline phase can be "locked in" by suitable heating.

Time-resolved pump-probe reflectivity measurements were carried out utilizing a near-infrared optical pulse of 20 fs duration and a center wavelength of 830 nm derived from a Ti:sapphire laser oscillator.
The linearly polarized pump (s-polarized) and probe (p-polarized) pulses were focused with a lens (f = 100 mm) onto the sample (held on a hot plate) at $ \simeq $ 45 and $ \simeq $ 50 degrees with respect to the sample normal, respectively.
The fluence of the pump pulse was kept below 100 $\mu$J/cm$^2$ so that the increase in the sample temperature due to laser irradiation could be neglected.
The transient reflectivity change of the probe pulse ($ \Delta $R/R) was recorded as a function of the pump-probe time delay.
The majority of the transient electronic response was subtracted by a band pass filter within the current amplifier.

\begin{figure}[htbp]
\includegraphics[width=60mm]{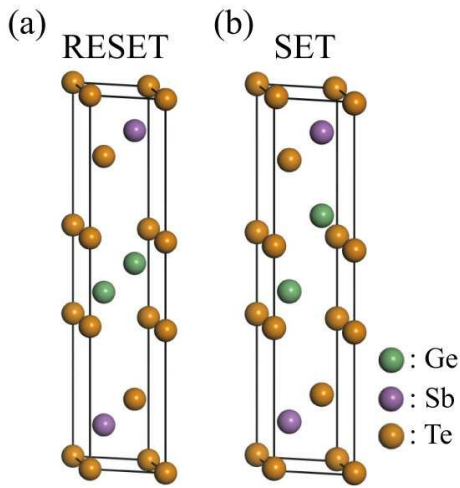}
\caption {Structural models of (a) the \{2,1\}-iPCM in the RESET and (b) SET states. The green, purple, and orange balls indicate Ge, Sb, and Te atoms, respectively.}
\label{Fig. 1}
\end{figure}

Figure 2 shows the $\Delta $R/R signals as a function of the pump-probe time delay for the two iPCM samples and the GST alloy sample obtained at 25 $^\circ $C and 180 $^\circ $C.
At 25 $^\circ $C, oscillations due to coherent phonons superimposed on the nonoscillatory electronic response were clearly observed in all samples.
The most noticeable difference between the iPCMs and the alloy is that the oscillations associated with coherent phonons survive even at 180 $^\circ $C in the iPCM samples while the amplitude of coherent phonon oscillations is strongly attenuated in the alloy sample at the same temperature.
After cooling the samples down to 25 $^\circ $C, further differences were observed (not shown) as follows.
In the alloy sample, the oscillation amplitude did not recover.
For the iPCM samples, on the contrary, the intensity of oscillation largely recovered.
The lifetime of the oscillation observed in \{2,4\}-iPCM sample was noticeably longer than that in the other samples as discussed later.

\begin{figure}[htbp]
\includegraphics[width=70mm]{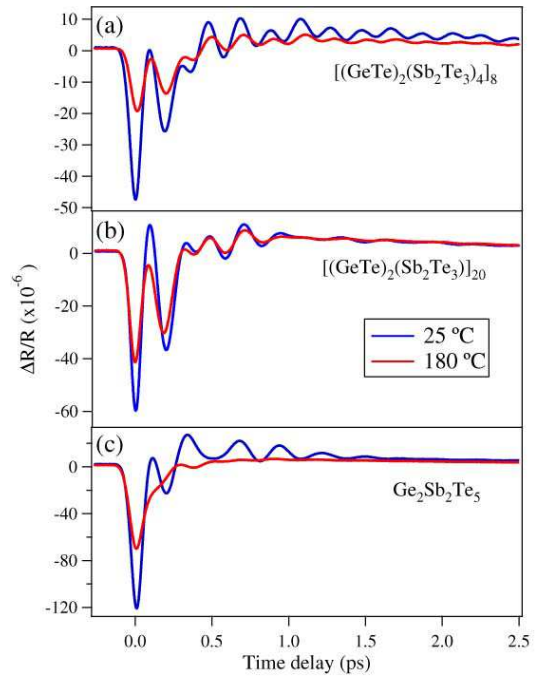}
\caption {Time-resolved $\Delta $R/R signals in (a) \{2,4\}-iPCM, (b) \{2,1\}-iPCM, and (c) GST alloy samples observed at 25 $^\circ $C and 180 $^\circ $C.}
\label{Fig. 2}
\end{figure}

Fourier transform (FT) spectra were obtained from the $\Delta $R/R signals for the iPCM samples and the GST alloy sample.
Figure 3 (a, b) shows FT spectra obtained for the \{2,4\}- and \{2,1\}-iPCM samples, respectively.
At 25 $^\circ $C, two dominant peaks were observed at around 3.4 THz (3.40 THz for \{2,4\}-iPCM and 3.36 THz for \{2,1\}-iPCM) and at around 5 THz (5.07 THz for \{2,4\}-iPCM and 4.82 THz for \{2,1\}-iPCM) in both samples.
Hereafter, for convenience, we refer to the two dominant peaks as the 3.4 THz mode and the 5 THz mode, respectively.
In addition, a small peak was observed at 2.11 THz only in the \{2,4\}-iPCM sample and a shoulder peak was seen at 3.78 THz in the \{2,1\}-iPCM sample.
Upon heating the iPCM samples above the phase transition temperature, the FT intensity drops but partially recovers after cooling to 25 $^\circ $C.
For the \{2,4\}-iPCM sample, the 5 THz peak shifted from 5.07 THz to 4.99 THz upon heating, and finally recovered to 5.07 THz after cooling.
The \{2,1\}-iPCM sample shows a similar yet more ambiguous trend due to the broad peak width and a small shift.
No significant shift of the other mode was confirmed. 

As shown in Fig. 3 (c), two main peaks were observed at 3.72 THz and 4.75 THz before heating the GST alloy.
The intensity of both peaks was drastically suppressed upon heating and the position of the phonon peaks become unclear due to the electronic-origin background superimposed on the phonon peaks.
Subsequent cooling increased the low frequency electronic background, yet the change in the phonon intensity was negligible.
In order to remove the background, the derivative of the $\Delta $R/R signal was taken and its FT spectra revealed that two main peaks appear at around 3.65 THz and 5.03 THz at 180 $^\circ $C (not shown.)
These results suggest that the amorphous alloy sample transformed into the fcc crystalline state upon heating and the crystalline state remain virtually unchanged upon cooling although there are some minor difference in the spectral shape from the previous report. \cite{Forst00}

\begin{figure}[htbp]
\includegraphics[width=70mm]{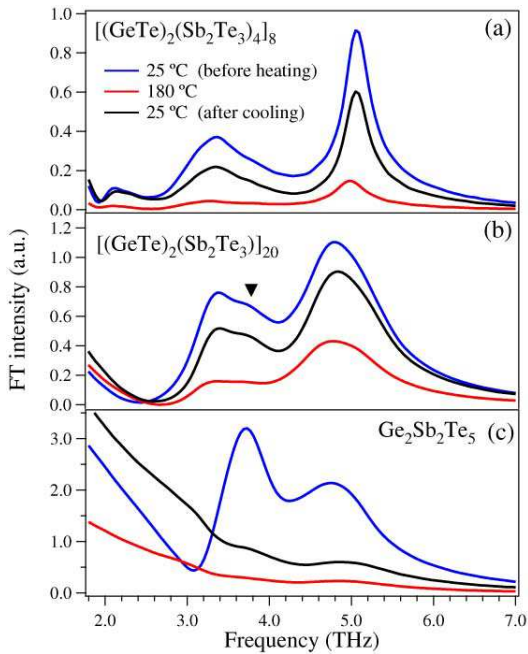}
\caption {FT spectra for (a) \{2,4\}-iPCM, (b) \{2,1\}-iPCM and (c) GST alloy films obtained from $\Delta $R/R data taken at 25 $^\circ $C before heating, at 180 $^\circ $C, and 25$^\circ $C after cooling. The black inverted triangle in (b) indicates the position of the shoulder peak at 3.78 THz.}
\label{Fig. 3}
\end{figure}

For the case of the GST alloy, the phase transition can be understood in the framework of a general amorphous-crystalline phase transition caused by thermal processes that include crystal growth for crystallization and a melt-quench for amorphization although several underlying mechanisms have been proposed.\cite{Kolobov04, Shportko08, Huang10}
In the current measurements, the amorphous GST alloy transformed into the fcc crystalline state after heating and the crystalline structure was preserved after cooling	.
In contrast, the phase transition mechanism in iPCMs is thought to be different from that in the alloy.
iPCM was named after its unique phase transition mechanism that arise at the interfaces between sub-layers.\cite{Simpson11}
Therefore, the clear contrast in the behavior of coherent phonons upon heating between iPCMs and GST alloy can be reasonably attributed to the difference in the phase transition mechanism.
The change in coherent phonon spectra for the iPCM films is much smaller than that for the GST alloy.
This result strongly suggests that the rearrangement of atoms during the phase transition is small in iPCMs compared to GST alloy as can be expected from the phase transition model.
In cooling the heated samples, a large part of the FT intensity recovers in the iPCM films but only a small change was observed in the GST alloy upon cooling as an irreversible phase transition occurred. \cite{Siegrist11}
Thus, the structural change in iPCMs associated with temperature on relatively slow time scales is reversible and consistent with the previous result of resistivity measurements.\cite{Bang14}
The incomplete recovery of the FT intensity can be explained by annealing effects which generally arise in PCM.
It should be noted that the thermal phase transition may be different from the electrical RESET-SET operation \cite{Simpson11} because the electrical phase transition is much faster (ns range) than the present heater-induced quasi-static
phase transition.
This is an area for further research.

Based on the previous reports on GST materials \cite{Forst00} and Bi$_{2}$Te$_{3}$/Sb$_{2}$Te$_{3}$ superlattice,\cite{Wang08} the $A_{1g}$ mode in Sb$_{2}$Te$_{3}$ sub-layers is a candidate for the lowest frequency mode found in the \{2,4\}-iPCM sample at 2.11 THz.
This mode also exists in the Sb$_{2}$Te$_{3}$ crystalline phase \cite{Richter77, Sosso09sbte} and hence the appearance of this mode suggests a bulk-like structure for the Sb$_{2}$Te$_{3}$ sub-layers in the \{2,4\}-iPCM sample.
Similarly, the 5 THz mode can be reasonably attributed to an additional $A_{1g}$ mode in the Sb$_{2}$Te$_{3}$ sub-layers.
With respect to remaining modes around 3.4 and 3.8 THz,
A$_{1}$ mode from the GeTe sub-layers are indicated in previous reports on other GST materials \cite{Forst00, Hase09, Hernandez-Rueda11, Shalini13}.
It is worth mentioning that even for an epitaxial Ge$_{2}$Sb$_{2}$Te$_{5}$ film fabricated on a GaSb(001) substrate, the symmetry of crystal is distorted to some extent due to intrinsically existing vacancies and it was found by extensive investigation of coherent phonons that the symmetry of phonon modes was significantly modified as a result of the imperfect crystal structure.\cite{Shalini13}
Polycrystalline samples can be expected to show more ambiguous phonon properties than an epitaxial sample.
 
\begin{figure}[htbp]
\includegraphics[width=80mm]{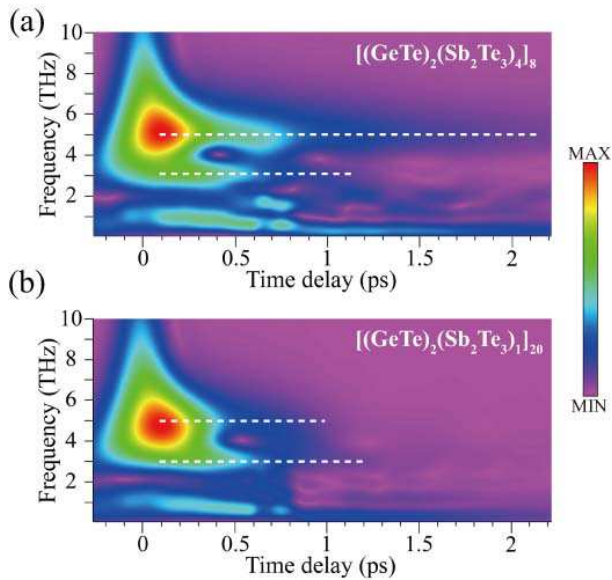}
\caption {Wavelet transform spectrograms for (a) \{2,4\}-iPCM and (b) \{2,1\}-iPCM calculated from $\Delta $R/R signals taken at 25 $^\circ $C before heating.
The white doted lines are guides for the eye.} 
\label{Fig. 4}
\end{figure}

Compared with the oscillatory pattern observed in the \{2,1\}-iPCM, the spectra of the \{2,4\}-iPCM possesses a long life time.
To clarify the decay time scale of the coherent phonons, we applied a wavelet transformation to carry out a time-frequency analysis \cite{Hase03} of the $\Delta $R/R signals measured on the pristine samples.
Figure 4 compares the wavelet spectrograms for the two iPCMs.
Here, the electronic background was subtracted by a combination of exponential functions.
As shown in Fig. 4 (a), the origin of the relatively long phonon lifetime in \{2,4\}-iPCM sample is found to be the 5 THz mode which assigned to the $A_{1g}$ mode in the Sb$_{2}$Te$_{3}$ sub-layers as mentioned above.
Since the \{2,4\}-iPCM sample has thicker Sb$_{2}$Te$_{3}$ sub-layers than the other samples, the structure of this layer should be more bulk-like as evidenced by the appearance of the 2.11 THz mode.
Hence we attributed the relatively long lifetime of the 5 THz mode to the small lattice deformation of the Sb$_{2}$Te$_{3}$ sub-layers with a small phonon scattering rate.
Note that the thicker Sb$_{2}$Te$_{3}$ sub-layers in the \{2,4\}-iPCM versus the \{2,1\}-iPCM are expected to serve to stabilize and orient the GeTe sub-layer and the current results is consistent with this supposition.
On the other hand, the 3.4 THz mode may be localized in the GeTe sub-layer or at the GeTe/Sb$_{2}$Te$_{3}$ interface because no Sb$_{2}$Te$_{3}$ sub-layer thickness dependence was found for this mode.
Also, another contributory factors such as scattering processes in the interfaces \cite{Wang08} and anharmonic phonon-phonon interactions \cite{Hase09} should also be taken into consideration.

In conclusion, we have carried out coherent phonon spectroscopy on [(GeTe)$_{2}$(Sb$_{2}$Te$_{3}$)$_{4}$]$_{8}$ and [(GeTe)$_{2}$(Sb$_{2}$Te$_{3}$)$_{1}$]$_{20}$ films which are the dominant structure for iPCM application and a conventional Ge$_{2}$Sb$_{2}$Te$_{5}$ alloy.
At 25 $^\circ $C, coherent phonon signals were clearly observed in all three samples.
By increasing the sample temperature up to 180 $^\circ $C, a large part of the coherent phonon signals were still observed in the iPCM samples although a RESET-to-SET phase transition was induced.
In contrast, a significant suppression of coherent phonons was found for the case of the alloy sample at the same temperature due to a phase transition from the amorphous state to the fcc crystalline state.
Upon subsequent cooling of the samples, it was found that the attenuation of coherent phonons introduced by heating was largely-reversible in iPCMs while that in GST alloy was irreversible.
These temperature-induced significant differences originate from the small atomic rearrangements associated with the RESET-SET phase transition process in iPCMs.
In addition, the results of a wavelet analysis applied to the $\Delta $R/R signals for two different iPCMs revealed the difference in the structure of Sb$_{2}$Te$_{3}$ sub-layers which is related to the orientation and stability of the iPCM structure. These features are directly relevant to the future high-performance PCM technology.

This work was supported in part by KAKENHI-26790063 from MEXT, Japan.
We acknowledge R. Kondou for sample preparation.

\end{document}